\documentclass{ieeetj}
\usepackage{cite}
\usepackage{amsmath,amssymb,amsfonts}
\usepackage{algorithmic}
\usepackage{graphicx,color}
\usepackage{textcomp}
\usepackage{xcolor}
\usepackage{hyperref}
\hypersetup{hidelinks=true}
\usepackage{algorithm,algorithmic}
\usepackage{comment}
\usepackage{subcaption}
\usepackage{float}
\def\BibTeX{{\rm B\kern-.05em{\sc i\kern-.025em b}\kern-.08em
    T\kern-.1667em\lower.7ex\hbox{E}\kern-.125emX}}
\AtBeginDocument{\definecolor{tmlcncolor}{cmyk}{0.93,0.59,0.15,0.02}\definecolor{NavyBlue}{RGB}{0,86,125}}
\newcommand{\review}[1]{{\textcolor{black}{#1}}}

\usepackage{threeparttable}
\usepackage{booktabs}
\usepackage{array}

\usepackage[table]{xcolor}

\def\OJlogo{\vspace{-4pt}$<$Society logo(s) and publication title will appear here.$>$}
\def\seclogo{\vspace{10pt}$<$Society logo(s) and publication title will appear here.$>$}

\def\authorrefmark#1{\ensuremath{^{\textbf{#1}}}}

\begin{document}
\receiveddate{XX Month, XXXX}
\reviseddate{XX Month, XXXX}
\accepteddate{XX Month, XXXX}
\publisheddate{XX Month, XXXX}
\currentdate{XX Month, XXXX}
\doiinfo{XXXX.2022.1234567}

\markboth{}{Author {et al.}}

\title{Towards Whole Hand and Wrist Kinematic Tracking
with a Wearable A-Mode Ultrasound Probe}

\author{Giusy Spacone\authorrefmark{1}, Graduate Student Member, IEEE, Luca Benini\authorrefmark{}\authorrefmark{2}, Fellow, IEEE\\ and  Andrea Cossettini\authorrefmark{1}, Senior Member, IEEE}
\affil{Integrated Systems Laboratory, ETH Zurich, Z{\"u}rich, Switzerland}
\affil{DEI, University of Bologna, Bologna, Italy }
\corresp{Corresponding author: Giusy Spacone (email: gspacone@iis.ee.ethz.ch).}
\authornote{{The authors acknowledge support from the ETH Research Grant \mbox{ETH-C-01-21-2} (Project ListenToLight)}}

\begin{abstract}
A-mode ultrasound (US) has emerged as a promising modality for hand and wrist motion tracking. Prior works have mainly addressed static gesture classification or regression of a few degrees of freedom (DoFs), typically relying on non-wearable systems and external computing devices, and highlight the need for strategies to ensure robustness to sensor repositioning. 
In this work, we propose a framework for robust whole-hand and wrist kinematic tracking via wearable A-mode US using the WULPUS platform, tackling the regression of 23 DoFs directly on the probe. First, we introduce a compact (11285 parameters) multi-output convolutional neural network combined with an incremental training strategy, which improves inter-session generalization and reduces mean absolute error by more than \review{17\%} compared to a non-incremental approach. Second, we demonstrate, for the first time, the feasibility of end-to-end hand and wrist kinematic tracking entirely on-device. We deploy the model on the WULPUS nRF52832 microcontroller, achieving $0.73~\text{mJ}$ per inference, $29.1~\text{ms}$ latency, and showing the feasibility of full operation (data acquisition, online inference, and BLE streaming of results) within \review{$33~\text{mW}$}, enabling up to \review{36} hours of continuous use and an 88\% reduction in wireless bandwidth compared to raw data transmission.

\end{abstract}


\maketitle
\vspace{-2cm}
\bstctlcite{IEEEexample:BSTcontrol}
\section{Introduction}

\IEEEPARstart{H}{and} gesture recognition (HGR) is a key enabler for next-generation human–machine interfaces \cite{guo_human-machine_2021}, with applications ranging from virtual reality interaction to prosthetic control and rehabilitation. While surface electromyography (sEMG) remains the predominant modality for HGR \cite{shin_hgrewview_2024}, A-mode ultrasound (US) is emerging as a promising alternative, as it probes deeper muscle layers and captures both coarse and fine motor activity \cite{us_neurorobotic_24}. Compared to conventional B-mode imaging, A-mode US is inherently more suitable for wearable systems, as it avoids the size, power, and computational overheads of image reconstruction \cite{zhou_2025_wearable}. 

Most prior works based on A-mode US have focused on the recognition of discrete hand gestures, with only a few studies addressing the continuous estimation of hand and wrist kinematics, a critical requirement for seamless control. Yang et al. \cite{yang_wearable_2021,yang_device_2021} demonstrated 2-DoF tracking using a custom 8-transducer system consuming $3.5~\text{W}$, achieving a cross-validated $R^2$ of 0.96. Sgambato et al. \cite{grandi_sgambato_high_2023,fournelle_portable_2021} used the MoUSE platform ($12~\text{W}$) with a 24-transducer armband to estimate 4 DoFs, reporting an $R^2$ of 0.94, which dropped below 0.8 when only four transducers were used. Spacone et al. \cite{spacone_tracking_24} reported the first results using a truly wearable, low-power acquisition system (WULPUS \cite{frey_wulpus_2022}), considering inter-session sensor repositioning for simultaneous tracking of 3 DoFs, achieving an inter-session RMSE of $11.11^{\circ}\pm4.14^{\circ}$ and a MAE of $8.46^{\circ}\pm 3.58^{\circ}$. More recently, Spacone et al. \cite{spacone_emgusfusion_25} extended the task to full-hand and wrist kinematics (23 DoFs), reporting inter-session errors of $9.3^{\circ}\pm1.8^{\circ}$ MAE, $13.1^{\circ}\pm2.6^{\circ}$ RMSE, and $R^2$ of $0.38\pm0.20$ on one subject. However, these works still face two critical limitations: accuracy degradation under sensor repositioning and the absence of edge deployment, leaving robust, embedded kinematic tracking as an open challenge. 

In this paper, we present two main contributions:

    \textit{1) Incremental fine-tuning for repositioning robustness:} We address accuracy degradation caused by sensor repositioning with a compact (11285 parameters) multi-output convolutional neural network and a user-centric incremental training strategy. This approach achieves $>$17\% reduction in MAE compared to non-incremental baselines, \review{while requiring only $\sim$5 minutes of new data to recalibrate after repositioning.}

    \textit{2) First end-to-end embedded deployment:} We demonstrate end-to-end 23 DoF kinematic tracking entirely on-device with WULPUS. Our deployment on the nRF52832 MCU achieves $0.73~\text{mJ}$ per inference and $29.1~\text{ms}$ latency, meeting the sub-100~ms requirements for prosthetic control \cite{farrell_2007_optimal}. The full pipeline (data acquisition, inference, and BLE transmission of results) consumes $<$\review{35}~mW, enabling over \review{36}~h of continuous operation on a 320~mAh Li-Po battery. Moreover, our approach reduces wireless bandwidth usage by 88\% compared to raw-data streaming, improving reliability under adverse network conditions and paving the way for scalable multi-node deployments.

\vspace{-0.2cm}
\section{Materials and Methods}
\subsection{Data collection protocol}

We implement the same data collection strategy as in \cite{spacone_emgusfusion_25}, extending it to two subjects, who provided informed consent. 
US data are collected with a WULPUS-based armband \cite{spacone_tracking_24} (see Fig.~\ref{fig:emb}), 
using four 32-element linear array transducers (Vermon) with a central frequency of $2.25~\text{MHz}$, equally spaced around the first half of the forearm. 
For each transducer, we use the 4 centermost channels shorted together as an equivalent single-channel element. 
Hydrogel pads \cite{hydrogel_pads} provide acoustic coupling.
WULPUS is operated with a pulse-repetition-rate of $30~\text{Hz}$, with one transducer activated at a time. 
The Manus Quantum Metaglove (MANUS, Netherlands) \cite{manus_quantum_glove} is used for ground-truth labelling, obtaining a stream of 20 finger joint angles and wrist orientation at $120~\text{Hz}$, synchronized to US data as described in \cite{spacone_emgusfusion_25}. 

In the data collection, the subject remains seated with the elbow supported at a 90$^\circ$ angle and performs a set of 11 hand and wrist movements, including hand opening–closing, pinches, thumb motions, and wrist flexion, extension, deviation, and supination.
We collect 3 sessions, including armband repositioning between sessions. Each session includes 5 sets. In each set, the 11 gestures are performed sequentially and repeated 6 times. Each gesture is held for 4 seconds with a 1-second rest between repetitions, resulting in approximately 5.5 minutes of recording per set. A 5-minute rest period between sessions prevents muscle fatigue.

\vspace{-0.2cm}
\subsection{Data Pre-Processing and Dataset Creation}
Wrist quaternions retrieved from the MANUS Core are used to compute wrist Euler angles (flexion/extension, radial–ulnar deviation, pronation–supination), assuming a ZYX rotation
sequence. A 1D Kalman filter with RTS smoothing is applied independently to the X, Y, and Z joint angles of Subject 2 to compensate for excessive noise in the GT angles (Subject 1’s signals did not require additional processing).
These angles are added to the finger joint angles, already in Euler form. MANUS data are downsampled to $7.5~\text{Hz}$, to match the time needed to complete one full arm scan (considering US acquisitions at $30~\text{Hz}$ and 4 transducers). 
Thus, each dataset sample contains four raw US frames
(dimension 400 × 4), along with the corresponding ground-truth labels (20 fingers and 3 wrist joint angles). 
\vspace{-0.5cm}
\subsection{Network architecture}
    The network architecture consists of two convolutional blocks (ks = [3 × 1]), each followed by rectified linear units activation function and max pooling ([4 × 1]), yielding a 16× spatial feature extraction and downsampling on a single channel. Features extracted from the channels are then flattened and followed by two fully connected layer heads (with 100 and 50 units, respectively) to regress the 20 hand and 3 wrist DoFs jointly. This lightweight design, adapted from the encoder–decoder architecture in \cite{spacone_emgusfusion_25}, is chosen to balance accuracy with embedded constraints of limited run-time memory ($<20~\text{kB}$ RAM, see below) and minimal inference time.
The architecture has 11285 parameters (60\% smaller than \cite{spacone_emgusfusion_25}) and is trained with a mean squared error (squared L2 norm) loss function.

\vspace{-0.5cm}
\subsection{Performance evaluation and incremental fine tuning}
We perform three types of analyses, evaluating the performance using the MAE, RMSE and the R\textsuperscript{2} score. 

\textit{(1)~Aggregated analysis}: we aggregate data from all acquisition sessions in a 5-fold cross-validation (CV) scheme: for each fold, three sets per session are used for training, one for validation, and one for testing. 
Models are trained for 50 epochs, with an early stop patience of 5 epochs, learning rate (l\textsubscript{r}) of $10^{-3}$, weight decay (w\textsubscript{d}) of $10^{-4}$ and using the Adam optimizer.

\textit{(2)~Inter-session analyses}: two sessions are used for training (the first four acquisition sets for each session for training and the remaining sets for validation). All sets from the unseen session are used for testing. We compute the performance for each set in the unseen session, as well as the average. We use the same training settings as in the aggregated experiment. This analysis reveals the impact of armband repositioning between sessions.

\textit{(3)~Incremental fine-tuning}: we train on two sessions (same as above) and then introduce a new acquisition session. Performance is first evaluated on the initial acquisition set in a zero-shot setting. We then fine-tune the model using 70\% of the data from the current set, with the remaining 30\% used for validation, and repeat this procedure for all sets. We cross-validate this approach across the three sessions. 

We use the same w\textsubscript{d} as in (1); for the l\textsubscript{r}, we ablate over four distinct values: $10^{-3}$, $2\times10^{-3}$, $10^{-4}$, $10^{-5}$. The maximum number of fine-tuning epochs is set to 25, with 5 epochs of patience.  We report the improvement achieved by fine-tuning in two cases: (i) comparing performance with and without fine-tuning across all acquisition sets, and (ii) comparing performance with and without fine-tuning when only the first acquisition set is used for calibration. 

\vspace{-0.2cm}
\subsection{Embedded implementation}
We deploy the network on the WULPUS nRF52832 MCU, which features $256~\text{kB}$ of non-volatile Flash and $64~\text{kB}$ of RAM memory (\review{$43~\text{kB}$ used for the acquisition firmware, with the remaining available for additional processing)}. 
The trained PyTorch models are converted through an intermediate ONNX representation into TensorFlow format, and subsequently into a TensorFlow Lite format \cite{tensorflow}. 
Post-training quantization is applied to the weights to convert the original 32-bit floating point (\texttt{float32}) representation to an 8-bit integer (\texttt{int8}) representation. From the resulting TensorFlow Lite FlatBuffer file (\texttt{.tflite}), the corresponding C and H files are generated, to be stored in the Flash memory of the target platform. The TensorFlow Lite Micro 
\cite{tflite_micro} library is used to port the model to the target platform.

\begin{figure}[h] 
  \centering
  \vspace{-0.1cm}
  \includegraphics[width=0.85\linewidth]{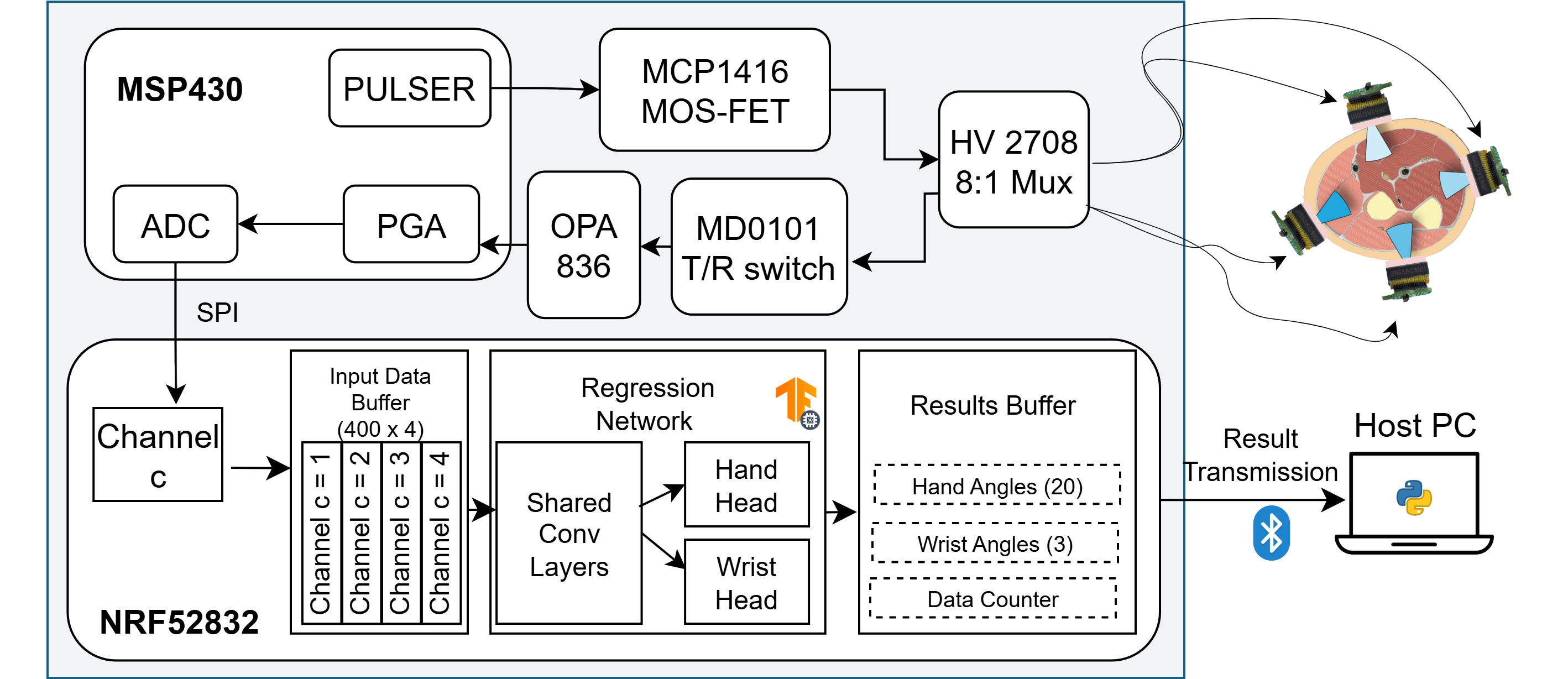}
  \vspace{-0.1cm}
  \caption{WULPUS architecture and detailed view of the data flow within the nRF52 MCU for embedded inference.}
  \vspace{-0.5cm}
  \label{fig:emb}
\end{figure}
Figure~\ref{fig:emb} illustrates the system data flow for the embedded implementation. Each newly acquired US frame (400 samples) is transferred via SPI transaction from the MSP430 to the nRF52 MCU and stored in a data buffer, containing the data collected from the four transducers. A regression pass can then be executed on each newly acquired transducer frame, without waiting for data from all four transducers. The predictions (92 bytes for the 23 output angles, stored as \texttt{float32}, plus 2 bytes for a packet counter ensuring data integrity), are transmitted via BLE to the host computer.
To evaluate the performance of the deployed network (energy and latency), we use the Nordic Power Profiler Kit II (PPK2, Nordic Semiconductor). 
The N6705B DC Power Analyzer (KeySight) 
is used for power measurements.

\vspace{-0.3cm}

\section{Results and Discussion}
\subsection{Tracking of Wrist-Hand Kinematics}
\textbf{1)~Aggregated analyses:} 
CV analyses for Subject 1 yield a MAE of $6.1^\circ\pm0.5^\circ$, RMSE of $8.6^\circ \pm 0.7^\circ$, and $R^2=0.69\pm0.03$. Subject 2 yields similar results, with a MAE of $6.4^\circ \pm 0.7^\circ$, RMSE of $9.1^\circ \pm 0.9^\circ$, and $R^2$ of $0.72\pm0.05$.
Results are comparable to our previous work \cite{spacone_emgusfusion_25} (RMSE $=8.5^\circ$, $R^2=0.7$), which also operates to 23 DoF, but with a much smaller network size (11285 vs. 27773 parameters), leading to lower FLASH, RAM, and MACCS. Compared to the results with four transducers on 4 DoF reported by \cite{grandi_sgambato_high_2023}, we achieve a lower R\textsuperscript{2} (0.69 vs 0.8). This is expected and attributed to the increased dimensionality of our problem ($6\times$ more DoF), and to the inclusion of fine-grained gestures (pinches, thumb movements), which increase the task complexity.

\textbf{2)~Inter-session analyses:} Subject 1 has a MAE of $9.5^\circ \pm 1.3^\circ$, RMSE of $13.2^\circ \pm 2.1^\circ$, and $R^2 = 0.34 \pm 0.22$. 
Subject 2 has a MAE of $11.2^\circ \pm 0.8^\circ$, RMSE of $16.0^\circ \pm 1.1^\circ$, and $R^2 = 0.27 \pm 0.04$.
Also in this case, the performance is comparable to \cite{spacone_emgusfusion_25}, with a much smaller network.

\begin{figure*}[]
  \centering
  \begin{subfigure}[t]{\textwidth}
    \centering
    \includegraphics[width=0.8\textwidth]{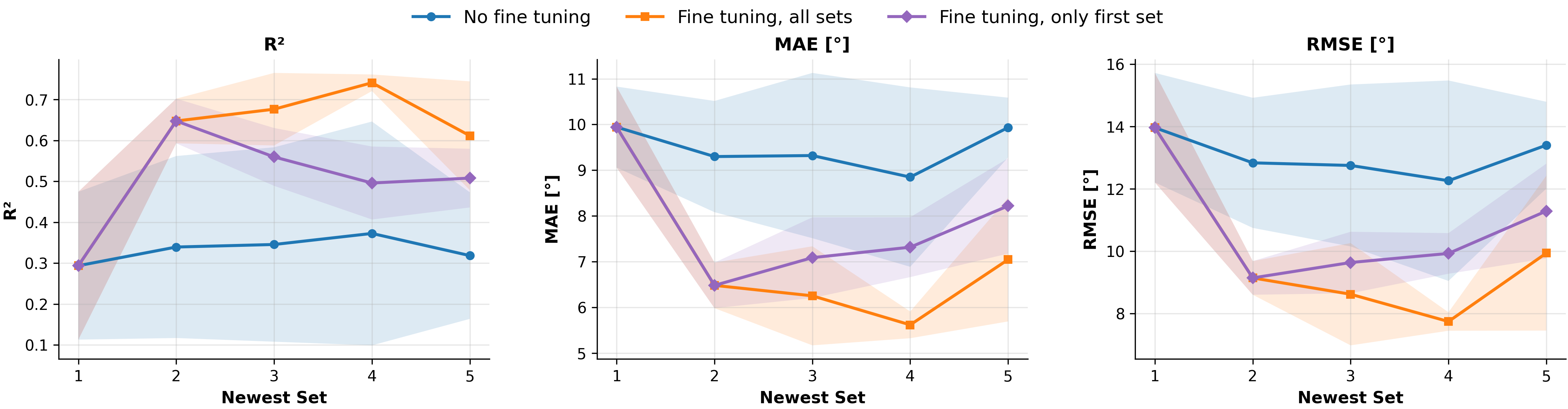}
    
    \label{fig:ft_s1}
  \end{subfigure} \hfill
  \vspace{-0.5cm}

  \caption{\review{Results for Subject 1 (CV performance across acquisition sets) for three cases: no fine tuning, progressive fine tuning, fine tuning with only one set after repositioning, }}
   \vspace{-0.6cm}
  \label{fig:ft}
\end{figure*}

\textbf{3)~Incremental fine tuning:} 
Fig.~\ref{fig:ft} shows the incremental fine-tuning results of Subject 1. Results are reported for a learning rate of $10^{-3}$, which yielded the best performance in the ablation studies.
For set 1, the \review{results with and without fine-tuning} are identical, as this set corresponds to the ``zero-shot" after repositioning (i.e., no fine-tuning has yet been applied). As additional data sets are progressively introduced, fine-tuning improves the performance. 
 
\begin{figure}[h] 
  \centering
  \includegraphics[width=0.95\linewidth]{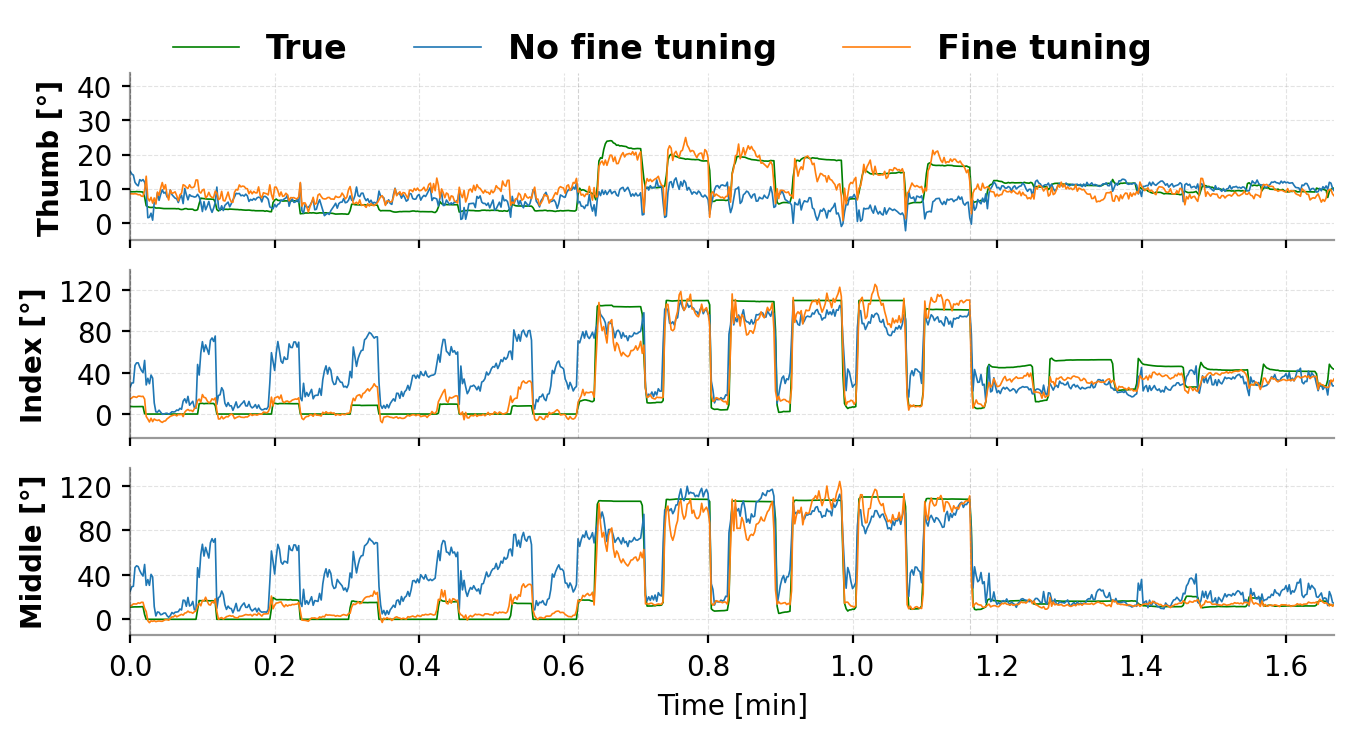}
  \vspace{-0.2cm}
  \caption{Example of a subset of ground-truth joint angles (green) versus model predictions (on the second acquisition set) without fine tuning (blue) and with fine tuning (orange).}
  \vspace{-0.8cm}
  \label{fig:results}
\end{figure}

Considering the average across all sets, the MAE for Subject 1 improves from $9.5^{\circ}$ (no fine-tuning) to $7.1^{\circ}$ (fine-tuning), corresponding to a 25\% error reduction. The $R^{2}$ also improves, from $0.33$ to $0.59$. Similarly, for Subject 2, the MAE improves from $11.2^{\circ}$ to $7.5^{\circ}$ (a 33\% error reduction) and the $R^{2}$ improves from $0.26$ to $0.61$. \review{When fine-tuning is performed only on the first new set of data, the average MAE for Subject 1 lowers to $7.8^{\circ}$ (17\% reduction) and the $R^{2}$ improves to $0.5$. For Subject 2, the MAE lowers to $8.7^{\circ}$ (22\% reduction) and the $R^{2}$ improves to $0.5$.} Fig. \ref{fig:results} shows an example of model prediction on the second set after only one set of fine-tuning for Subject 2. Improvements in the metrics, particularly in the $R^{2}$ score, are predominantly associated with a reduction in the prediction offset. \review{We attribute the performance degradation across acquisition sets when fine-tuning is performed only on the first set to small transducer displacements, which may have occurred during rest phases between sets or during movement execution.} 
\review{Since acquiring a new set requires only 5.5 minutes of data, this demonstrates that on-demand recalibration with A-mode US can be performed with minimal effort}. While such strategies are already well established for sEMG and are widely integrated into commercial myoelectric prostheses \cite{cambell_coadaptation_2025}, only one prior study has investigated incremental fine-tuning with US for gesture classification using B-mode imaging \cite{rothenber_incrementalBmode_2024}, and our work demonstrates its first application to A-mode.

\vspace{-0.55cm}
\vspace{-0.5cm}
\subsection{Embedded Implementation}
The deployed network architecture has a total memory footprint of $16.8~\text{kB}$, corresponding to only $7\%$ of the total available Flash memory. The RAM memory allocation is $10.1~\text{kB}$, corresponding to $16\%$ of the available memory. The time to produce one output is $29.1~\text{ms}$, meeting the latency requirements of embedded applications such as prosthetic control \cite{farrell_2007_optimal}, with an energy per inference of $0.73~\text{mJ}$. 
The power consumption of the end-to-end system, required to retrieve one US frame, produce one output, and transmit the results via BLE is \review{33}$~\text{mW}$. 
Of these, 23$~\text{mW}$ are required by the nRF52 to retrieve data from the MSP MCU, perform one computation, and transmit the output.
Transmitting only the prediction instead of raw data offers $88\%$ wireless link bandwidth improvements.

 \vspace{-0.4cm}
\section{Conclusion}
We presented a framework for whole-hand and wrist kinematic tracking using a fully dry, WULPUS-based armband with four ultrasound transducers and edge-computing. 

We employed a compact (11285 parameters) multi-task convolutional neural network to regress 20 hand and 3 wrist degrees of freedom, considering inter-session armband removal to mimic realistic usage conditions. To address the performance drop caused by sensor repositioning, we introduced a user-centric incremental fine-tuning approach, which reduced the average MAE by \review{more than 17\%} compared to a non-incremental strategy \review{with only $\sim$5 minutes of new data for recalibration.} 
To the best of our knowledge, this was the first demonstration of on-demand fine-tuning for A-mode systems, suggesting a viable path for integration into future commercial prosthetic devices to enable robust, ultrasound-based control schemes.

We further deployed our network architecture on the WULPUS nRF52832 microcontroller. With an energy cost of only $0.73~\text{mJ}$ per inference and a computation time of $29.1~\text{ms}$, our approach satisfies the constraints of embedded applications such as prosthetic control. We demonstrated the feasibility of end-to-end data collection, inference, and results streaming, achieving an 88\% reduction in wireless bandwidth compared to raw data transmission, thereby improving system reliability under adverse network conditions.

Although the study was conducted on two subjects, it demonstrated the feasibility of incremental calibration and embedded deployment. Future work will extend validation to a larger cohort and enhance the acquisition setup by increasing the number of ultrasound transducers to improve spatial coverage and robustness.

\vspace{-0.3cm}
\section*{Acknowledgment}
We thank S. Frey, V. Kartsch, C. Leitner, L. Lamberti (ETH Z{\"u}rich) and M. Orlandi (University of Bologna) for technical support and fruitful discussions. 

\bibliographystyle{IEEEtran}
\vspace{-0.2cm}
\bibliography{bib}

@IEEEtranBSTCTL{IEEEexample:BSTcontrol,
  CTLuse_forced_etal = "yes",
  CTLmax_names_forced_etal = "2",
  CTLnames_show_etal = "1"
}

@article{guo_human-machine_2021,
	title = {Human-Machine Interaction Sensing Technology Based on Hand Gesture Recognition: A Review},
    year = 2021,
	issn = {2168-2305},
	doi = {10.1109/THMS.2021.3086003},
	shorttitle = {Human-Machine Interaction Sensing Technology Based on Hand Gesture Recognition}, 
    volume={51},
	pages = {300--309},
	number = {4},
	journal = {{IEEE} Trans. Hum.-Mach. Syst.},
	author = {Guo, Lin and Lu, Zongxing and Yao, Ligang},
	urldate = {2025-05-17},
	date = {2021-08},
}

@ARTICLE{shin_hgrewview_2024,
  author={Shin, Jungpil and Miah, Abu Saleh Musa and Kabir, Md. Humaun and Rahim, Md. Abdur and Al Shiam, Abdullah},
  journal={IEEE Access}, 
  title={A Methodological and Structural Review of Hand Gesture Recognition Across Diverse Data Modalities}, 
  year={2024},
volume={12},
  pages={142606--142639},
  doi={10.1109/ACCESS.2024.3456436},
}

@article{zhou_2025_wearable,
  title={Wearable ultrasound technology},
  author={Zhou, Sai and Park, Geonho and Lin, Muyang and Yang, Xinyi and Xu, Sheng},
  journal={Nat. Rev. Bioeng.},
  volume={3},
  number={10},
  pages={835--854},
  year={2025},
  publisher={Nature Publishing Group UK London}
  
}

@ARTICLE{us_neurorobotic_24,
  author={Yang, Xingchen and Castellini, Claudio and Farina, Dario and Liu, Honghai},
  journal={IEEE Trans. Syst. Man Cybern.: Syst.}, 
  title={Ultrasound as a Neurorobotic Interface: A Review}, 
  year={2024},
  volume={54},
  number={6},
  pages={3534-3546},
  doi={10.1109/TSMC.2024.3358960}}

@ARTICLE{spacone_tracking_24,
  author={Spacone, Giusy and others},
  journal={IEEE Trans. Biomed. Circuits Syst.}, 
  title={Tracking of Wrist and Hand Kinematics With Ultra Low Power Wearable A-Mode Ultrasound}, 
  year={2025},
  volume={19},
  number={3},
  pages={536-548},
  doi={10.1109/TBCAS.2024.3465239}}

@inproceedings{spacone_emgusfusion_25,
  title={Wearable and Ultra-Low-Power Fusion of EMG and A-Mode US for Hand-Wrist Kinematic Tracking},
  author={Spacone, Giusy and Frey, Sebastian and Orlandi, Mattia and Rapa, Pierangelo Maria and Kartsch, Victor and Benatti, Simone and Benini, Luca and Cossettini, Andrea},
  booktitle={{2025 IEEE BioCAS}},
  pages={{314--318}},
  year={{2025}},
}

@article{yang_wearable_2021,
	title = {Wearable Ultrasound-Based Decoding of Simultaneous Wrist/Hand Kinematics},
	volume = {68},
	issn = {1557-9948},
	doi = {10.1109/TIE.2020.3020037},
	pages = {8667--8675},
	number = {9},
	   journal = {{IEEE} Trans. Ind. Electron.}, 
	author = {Yang, Xingchen and Zhou, Yu and Liu, Honghai},
	urldate = {2024-04-06},
	date = {2021-09},
year = 2021
}

@article{yang_device_2021,
  author={Yang, Xingchen and others},
journal={IEEE Trans. Syst. Man Cybern.: Syst.},
  title={A Wearable Ultrasound System for Sensing Muscular Morphological Deformations}, 
  year={2021},
  volume={51},
  number={6},
  pages={3370-3379},
  doi={10.1109/TSMC.2019.2924984}}

@article{grandi_sgambato_high_2023,

  author={Sgambato, Bruno Grandi and others},
journal={IEEE Trans. Biomed. Eng.}, 


  title={High Performance Wearable Ultrasound as a Human-Machine Interface for Wrist and Hand Kinematic Tracking}, 

  year={2024},

  volume={71},

  number={2},

  pages={484-493},

  doi={10.1109/TBME.2023.3307952}}

@article{fournelle_portable_2021,
	title = {Portable Ultrasound Research System for Use in Automated Bladder Monitoring with Machine-Learning-Based Segmentation},
	volume = {21},
	rights = {http://creativecommons.org/licenses/by/3.0/},
	issn = {1424-8220},
	doi = {10.3390/s21196481},
	pages = {6481},
	number = {19},
	journal = {Sensors},
	author = {Fournelle, Marc and others},
	urldate = {2024-03-03},
	date = {2021-01},
	langid = {english},
year=2021
}

@article{farrell_2007_optimal,
  title={The optimal controller delay for myoelectric prostheses},
  author={Farrell, Todd and others},
  journal={IEEE Trans. Neural Syst. Rehabil. Eng.},
  volume={15},
  number={1},
  pages={111--118},
  year={2007},
  publisher={IEEE}
}

@ARTICLE{cambell_coadaptation_2025,
  author={Campbell, Evan and others},
  
  journal={IEEE Trans. Neural Syst. Rehabil. Eng.},
  title={(Un)supervised (Co)adaptation via Incremental Learning for Myoelectric Control: Motivation, Review, and Future Directions}, 
  year={2025},
  volume={33},
  number={},
  pages={3565-3582},
  doi={10.1109/TNSRE.2025.3602397}}

@inproceedings{rothenber_incrementalBmode_2024,
  title={Improving Intersession Reproducibility for Forearm Ultrasound based Hand Gesture Classification through an Incremental Learning Approach},
  author={Rothenberg, Jack and others},
  booktitle={2024 IEEE UFFC-JS},
  pages={1--4},
  year={2024}
}

@INPROCEEDINGS{frey_wulpus_2022,
  author={Frey, Sebastian and others},
  booktitle={2022 IEEE IUS}, 
  title={WULPUS: a Wearable Ultra Low-Power Ultrasound probe for multi-day monitoring of carotid artery and muscle activity}, 
  year={2022},
  volume={},
  number={},
  pages={1-4},
  doi={10.1109/IUS54386.2022.9958156}}

@online{hydrogel_pads, 
  title = {Hydrogel Pad}, 
  note = {{https://amzn.eu/d/9dQ3nRG}}, 
  urldate = {2025-05-02},
}

@online{manus_quantum_glove,
  author       = {{MANUS Technology Group}},
  title        = {{Quantum Metagloves}},
  year         = {2025},
note = {{https://www.manus-meta.com/products/quantum-metagloves}}

}

@online{tensorflow,
  title = {{TensorFlow}},
  note = {{https://github.com/tensorflow/tensorflow}}
}

@online{tflite_micro,
  title = {{TF Lite Micro}},
  note = {{https://github.com/tensorflow/tflite-micro}}
}

\end{document}